\documentclass[aps,prc,showpacs,twocolumn,superscriptaddress]{revtex4}
\usepackage{multirow}
\usepackage{graphicx}
\usepackage{color}
\usepackage{amsmath}
\usepackage{amssymb}
\usepackage[colorlinks=true]{hyperref}
\usepackage{tikz}
\usepackage{etoolbox}

\usepackage[utf8]{inputenc}

\newcommand {\snn}      {\sqrt{s_{_{\rm NN}}}}

\newcommand {\hijing}   {{\tt HIJING}}
\newcommand {\urqmd}      {{\tt UrQMD}}
\newcommand {\vishnu}   {{\tt iEBE-VISHNU}}
\newcommand {\vish}     {{\tt VISHNU}}
\newcommand {\vis}      {{\tt VISH2+1}}

\newcommand {\atlas}    {{\small ATLAS}}

\definecolor{dgreen}{cmyk}{1.,0.,1.,0.4}        
\definecolor{orange}{cmyk}{0.,0.353,1.,0.}    

\begin{document}
\title{
\begin{flushright}
{
}
\end{flushright}
Hydrodynamic Collectivity in  Proton--Proton Collisions at 13 TeV}

\author{Wenbin Zhao}
\affiliation{Department of Physics and State Key Laboratory of Nuclear Physics and Technology, Peking University, Beijing 100871, China}
\affiliation{Collaborative Innovation Center of Quantum Matter, Beijing 100871, China}

\author{You Zhou}
\email{You Zhou: you.zhou@cern.ch}
\affiliation{Niels Bohr Institute, University of Copenhagen, Blegdamsvej 17, 2100 Copenhagen, Denmark}

\author{Haojie Xu}
\affiliation{School of Science, Huzhou University, Huzhou 313000, China}
\affiliation{Department of Physics and State Key Laboratory of Nuclear Physics and Technology, Peking University, Beijing 100871, China}

\author{Weitian Deng}
\affiliation{School of Physics, Huazhong University of Science and Technology, Wuhan 430074, China}

\author{Huichao Song}
\email{Huichao Song: huichaosong@pku.edu.cn}
\affiliation{Department of Physics and State Key Laboratory of Nuclear Physics and Technology, Peking University, Beijing 100871, China}
\affiliation{Collaborative Innovation Center of Quantum Matter, Beijing 100871, China}
\affiliation{Center for High Energy Physics, Peking University, Beijing 100871, China}

\date{\today}

\begin{abstract}
In this paper, we investigate the hydrodynamic collectivity in proton--proton (p--p) collisions at 13 TeV, using \vishnu\ hybrid model with \hijing\ initial conditions. With properly tuned parameters, our model simulations can remarkably describe all the measured 2-particle correlations, including integrated and differential elliptic flow coefficients for all charged and identified hadrons ($K_S^0$, $\Lambda$). However, our model calculations show positive 4-particle cumulant $c_{2}\{4\}$ in high multiplicity pp collisions, and can not reproduce the negative $c_{2}\{4\}$ measured in experiment. Further investigations on the \hijing\ initial conditions show that the fluctuations of the second order anisotropy coefficient $\varepsilon_{2}$ increases with the increase of its mean value, which leads to a similar trend of the flow fluctuations. For a simultaneous description of the 2- and 4- particle cumulants within the hydrodynamic framework, it is required to have significant improvements on initial condition for pp collisions, which is still lacking of knowledge at the moment.

\end{abstract}


\pacs{25.75.Ld, 25.75.Gz}

\maketitle

\clearpage

\section{Introduction}
\label{section1}

One of the main goal of the heavy-ion program at Relativistic Heavy Ion Collider (RHIC) and the Large Hadron Collider (LHC) is to create a novel state of matter, the Quark-Gluon Plasma (QGP), and study its properties. The anisotropic flow, that evaluates the anisotropy of the momentum distribution of final produced particles, is sensitive to both initial state fluctuations and the QGP transport properties~\cite{Voloshin:2008dg,Heinz:2013th,Gale:2013da,Luzum:2013yya,Jia:2014jca,Song:2013gia,Song:2017wtw}. Fruitful flow data~\cite{ALICE:2011ab,Abelev:2014pua,Adam:2016izf,ALICE:2016kpq,Acharya:2017ino,Acharya:2017zfg,Acharya:2017gsw,CMS:2013bza,Sirunyan:2017fts,Chatrchyan:2013kba,ATLAS:2011ah,ATLAS:2012at,Aad:2013xma,Aad:2014fla,Aad:2015lwa,Aaboud:2017tql} and the successfully descriptions by hydrodynamic calculations~\cite{Song:2010mg,Niemi:2011ix,Schenke:2010rr,Song:2012ua,Gale:2012rq,Bernhard:2016tnd, McDonald:2016vlt,Zhao:2017yhj}, reveale that the created QGP fireball behaves like a nearly perfect liquid with a very small specific shear viscosity $\eta/s$ close to the conjectured lowest bound $1/4\pi$~\cite{Kovtun:2004de}.

The high energy proton--lead (p--Pb) and proton--proton (p--p) collisions at the LHC were originally aimed to provide the reference data for the high energy nucleus-nucleus collisions. However, various unexpected phenomena have been observed in these small systems, especially in the high multiplicity region. One surprising discovery is the long-range ``ridge'' structures in two-particle azimuthal correlations with a large pseudo-rapidity separation in high multiplicity p--Pb and p--p collisions~\cite{Khachatryan:2010gv,Li:2012hc,Khachatryan:2015lva,Aad:2015gqa,Khachatryan:2016txc}. Such long-range correlation structures were firstly discovered in Au--Au and Pb--Pb collisions and interpreted as a signature of the collective expansion. In general, the theoretical interpolations for the long-range ``ridge'' structure in the small systems
can be classified into three big categories: final state interactions, such as hydrodynamic expansion~\cite{Bozek:2011if,Bozek:2013ska,Bzdak:2013zma,Qin:2013bha,Weller:2017tsr,Schenke:2014zha}, parton cascade~\cite{Bzdak:2014dia,Ma:2014pva,Bozek:2015swa,Li:2016ubw}, hadronic rescattering~\cite{Zhou:2015iba}, rope and shoving mechanism~\cite{Bierlich:2017vhg}, initial state effects related to the gluon saturation~\cite{Dusling:2012iga,Dusling:2012cg,
Dusling:2013qoz,Dusling:2014oha,Dumitru:2014dra,Dumitru:2014vka,Noronha:2014vva,Schenke:2016lrs} and combinations of both initial and final state effects~\cite{Mantysaari:2017cni}. For recent theoretical progresses, please refer to~\cite{Song:2017wtw,Dusling:2015gta}.

In experiments, one of the crucial questions on the ``ridge'' structure is whether it arises from correlations of all particles related to  the collective flow or it only involves with the correlations from few particles, e.g. from resonance decays or jets, which is defined as non-flow. In small pp and p--Pb systems, the non-flow contributions are always significant, even in case that collective expansion has been developed. It is thus necessary to remove such non-flow effects before comparing the data with the model calculations. Based on different assumptions, various non-flow subtraction methods, e.g. template fit~\cite{Aad:2015gqa,Aaboud:2016yar,ATLAS:2017tqk} and peripheral subtraction~\cite{Khachatryan:2016txc} have been applied to the measurements of 2-particle correlations in pp collisions at 13 TeV, which yield different non-flow subtracted results. Currently, it is still unclear which one is a better approach to remove the non-flow effects.

Compared with the 2-particle correlations, multi-particle cumulants are less influenced by the non-flow effects, which are expected as one of the key observables to evaluate the anisotropic collectivity of the small systems. Besides, the multi-particle and 2-particle cumulants show different sensitivities to event-by-event flow fluctuations~\cite{Voloshin:2008dg,Borghini:2000sa,Jia:2014pza}. An extensive measurement of these different cumulants could provide tight constraints on the initial state fluctuations.

In order to extract real values of the flow coefficients, the  2-, 4-, 6- and 8-particle cumulants are expected to carry positive, negative, positive and negative signs, respectively. Such ``changing sign pattern'' has been observed in the measured 2- and multi-particle cumulants in Pb--Pb collisions at the LHC, where the created QGP fireballs undergo fast collective expansion~\cite{ALICE:2011ab, Adam:2016izf}. Based on the similar idea, it is proposed to measure the 2- and multi-particle cumulants to evaluate the collectivity in high multiplicity pp collisions at 13 TeV. However, it was found that the standard multi-particle cumulants in the small systems are still largely affected  by the residual non-flow and its fluctuations, which even presents fake flow signals with the ``right sign''. Recently, Ref.~\cite{Jia:2017hbm, Huo:2017nms}
developed 2- and 3-subevent methods for the multi-particle cumulants, which further remove the residual non-flow from jet.
The related measurements from ATLAS have confirmed the observations of positive 2-particle cumulants and negative 4-particle cumulants in high multiplicity pp collisions at 13 TeV~\cite{Aaboud:2017blb}. It is thus the proper time to study and evaluate these possible collective flow signal, using the hydrodynamic calculations. In this paper, we will implement \vishnu\ hybrid model with {\tt HIJING} initial conditions to study 2- and 4-particle correlations in pp collisions at $\snn = 13$ TeV, together with a detailed examinations of the initial state fluctuations from {\tt HIJING}.

\section{The model and set-ups}
\label{sec:model}

\vishnu\ ~\cite{Shen:2014vra} is an event-by-event simulation version of the early developed hybrid model \vish~\cite{Song:2010aq} that combines (2+1)-d viscous hydrodynamics \vis~\cite{Song:2007fn,Song:2007ux} to describe the QGP expansion with a hadron cascades model \urqmd~\cite{Bass:1998ca,Bleicher:1999xi} to simulate the evolution of hadronic matter. In the viscous hydrodynamics part,  \vis\ solves the transport equations for the energy-momentum tensor $T^{\mu \nu}$ and shear stress tensor $\pi^{\mu \nu}$ with a state-of-art equation of state (EoS) s95-PCE~\cite{Bernhard:2016tnd,Bazavov:2014pvz} as an input. For simplicity, we neglect the bulk viscosity, net baryon density and heat conductivity and
assume that specific shear viscosity $\eta/s$ is a constant. The hydrodynamic evolution matches
the hadron cascade simulations at a switching temperature $T_{\rm sw}$, where various hadrons are emitted from the
switching hyper-surface for the succeeding \urqmd\ evolution.

In our calculations,  we use the modified \hijing\ model~\cite{Xu:2012au} to generate fluctuating initial profiles for the succeeding \vishnu\ simulations in high energy pp collisions. In \hijing~\cite{Wang:1991hta, Deng:2010mv, Deng:2010xg}, the produced jet pairs and excited nucleus are treated as independent strings, where the hard jet productions are calculated by  pQCD, and the soft interactions are treated as gluon exchange within Lund string model. Here, we assume these strings break into partons independently and quickly form several hot spots for the succeeding hydrodynamic evolution.
Following~\cite{Deng:2011at}, the center positions of the strings $(x_c, y_c)$ are sampled by the Saxon-Woods distribution and the positions of the produced partons within a string are sampled with a Gaussian distribution
$\exp\left.(-\frac{(x-x_{c})^{2}+(y-y_{c})^{2}}{2\sigma_{R}^2}\right.)$, where $\sigma_{R}$ is the Gaussian smearing factor of the string.

\begin{table}[tbp]
\centering  %
\caption{Four sets parameters used in \vishnu\ simulations  with \hijing\ initial conditions for pp collisions at 13 TeV. }
         \label{tb:para}
\begin{tabular}{|l|c|c|c|c|c|c|}
\hline
   &$\sigma_R$&$\sigma_{0}$&$\tau_0$ & $\eta/s$ & K &$T_{\rm sw}$(MeV) \\ \hline  %
Para-I  &1.0&0.4&0.1&0.07 &1.26 &147\\        \hline  %
Para-II &0.8&0.4&0.2&0.08 &1.25 &148\\        \hline  %
Para-III &0.4&0.2&0.6&0.20 &1.13  &148\\        \hline  %
Para-IV &0.6&0.4&0.4&0.05 &1.28  &147\\        \hline  %
\end{tabular}
\end{table}

Following \cite{Xu:2016hmp}, the initial energy density profiles in the transverse plane are constructed from the energy decompositions of
emitted partons of \hijing\, together
 with an additional Gaussian smearing\cite{Xu:2016hmp}
\begin{equation}
	\epsilon(x,y) = K\sum_{i}\frac{p_{i}U_{0}}{2\pi\sigma_0^{2}\tau_{0}\Delta\eta_{s}}\exp\left.(-\frac{(x-x_{i})^{2}+(y-y_{i})^{2}}{2\sigma_0^2}\right.), \label{eq:epsilon}
\end{equation}
where $\sigma_0$ is the Gaussian smearing factor, $p_{i}$ is the momentum of the produced parton $i$
and $K$ is an additional normalization factor, $U_0$ is the initial flow velocity of the corresponding fluid cell.
Here, we assume zero transverse initial flow and only consider the  partons within the spacial mid-rapidity  $|\eta_s|<1$ (for related details, please refer to~\cite{Xu:2016hmp}).

\begin{figure}[t]
\begin{center}
\includegraphics[width=0.42\textwidth]{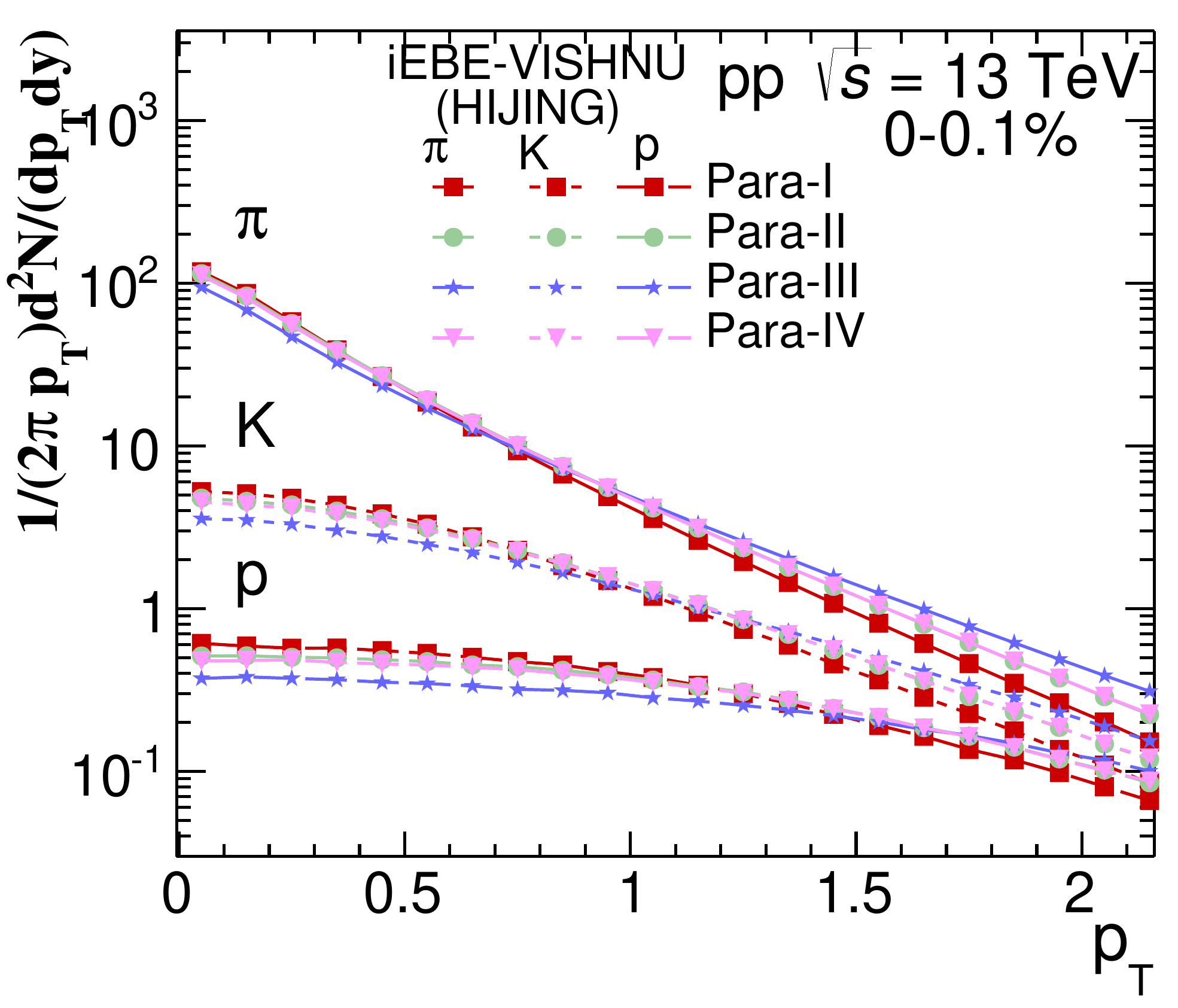}
\caption{(Color online) The $p_{\rm T}$-spectra of pions, kaons and protons in 0-0.1$\%$ pp collisions at 13 TeV, calculated by \vishnu\  hybrid model with \hijing\ initial conditions. }
\label{fig:pt_spectra}
\end{center}
\end{figure}

In \vishnu\ simulations with \hijing\ initial conditions, the hydrodynamic starting time $\tau_0$, switching temperature $T_{\rm sw}$, specific shear viscosity $\eta/s$, the Gaussian smearing width $\sigma_R$ and $\sigma_0$, and normalization factor $K$ are free parameters, which need to be fixed by experimental data. In general, one uses total multiplicity, $p_T$ spectra of identified hadrons and integrated flow harmonics of all charged hadrons to tune the related parameters in the hydrodynamic calculations~\cite{Xu:2016hmp,Zhu:2016puf,Zhao:2017yhj}. However, not all these needed data are available in pp collisions at 13 TeV. For example, the $p_{\rm T}$ spectra has not been measured and released. Here, we assume that slope of $p_{\rm T}$ spectra do not significantly change from 7 TeV to 13 TeV, and use these slopes of pions and protons at 7 TeV~\cite{DerradideSouza:2016kfn}, the total multiplicity~\cite{Khachatryan:2016txc} and $v_2\{2\}$~\cite{Khachatryan:2016txc, ATLAS:2017tqk} at 13 TeV to partially constrain the parameters in \hijing\ and \vishnu. In fact, these available data are not enough to fully fix  all these parameters in our model calculations, especially considering that the measured $v_2\{2\}$ from CMS~\cite{Khachatryan:2016txc} and ATLAS ~\cite{Aaboud:2016yar,ATLAS:2017tqk} differ by 20\%. We thus select four possible parameter sets as listed in Tab.~\ref{tb:para}, that roughly fit slope of $p_{\rm T}$ spectra at 7 TeV and fit the measured $v_2\{2\}$ from either ATLAS or CMS collaborations, for the following calculations and investigations in Sec.~III. The predicted $p_{\rm T}$ spectra of pions, kaons and protons in 0- 0.1\% pp collisions at 13 TeV  are shown in Fig.~\ref{fig:pt_spectra}, where the centralities are cut by the multiplicity of all charged hadrons at mid-rapidity $| \eta|<2.4$~\footnote{In our {\tt iEBE-VISHNU} simulations, the particle event generator between 2+1-d hydrodynamics and {\tt UrQMD} samples particles within the the momentum rapidity range $|y|<3.0$.  After the UrQMD evolution, the boost-invariance are approximately kept within $|y|<2.4$.}. The spectra at 13 TeV have similar slopes as the ones at 7 TeV, which can be measured in experiments
in the near future.


\section{Results and Discussion}
\label{sec:result}
The 2- and multi-particle correlations are common measurements to study anisotropic azimuthal correlations, which can be calculated with the Q-cumulant method~\cite{Bilandzic:2010jr} and the Generic Framework~\cite{Zhou:2015iba}. In our calculations,
these two methods are identical, since it does not involve any inefficiency in azimuthal acceptance or tracking efficiency as happened in experiments. We thus follow the same procedure as in our early study~\cite{Zhou:2015iba} to calculate the
2- and multi-particle correlations as well as the related flow harmonics.

\subsection{2-particle cumulant}

  \begin{figure}[thb]
\begin{center}
\includegraphics[width=0.49\textwidth]{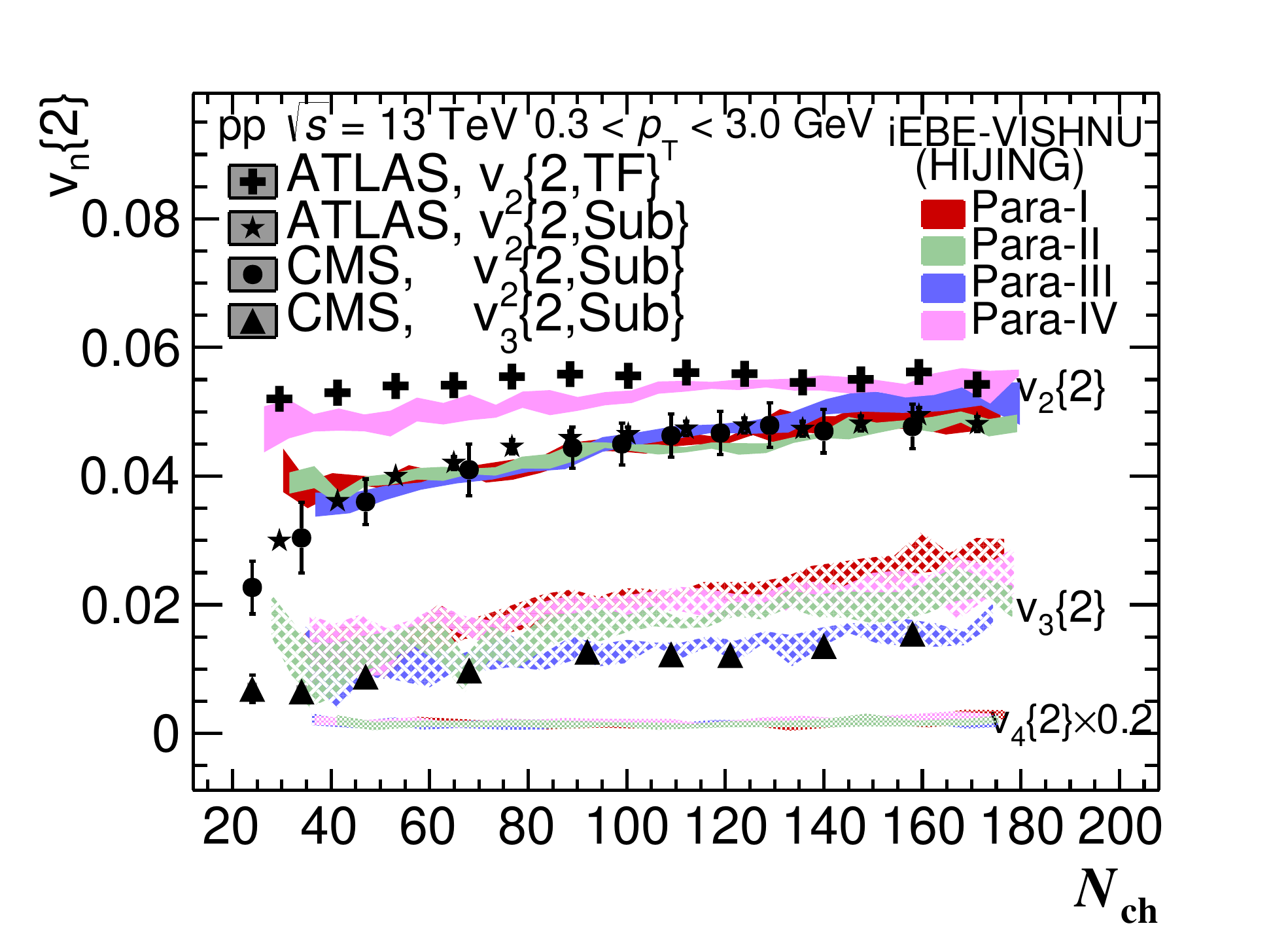}
\caption{(Color online)  $v_2\{2\}$, $v_3\{2\}$ and $v_4\{2\}$ in  pp collisions at 13 TeV, calculated by \vishnu\ with \hijing\ initial conditions. The CMS and ATLAS  data are taken from~\cite{Khachatryan:2016txc} and ~\cite{ATLAS:2017tqk}, respectively.}
\label{fig:v22}
\end{center}
\end{figure}

\begin{figure*}[thb]
\begin{center}
\includegraphics[width=0.95\textwidth]{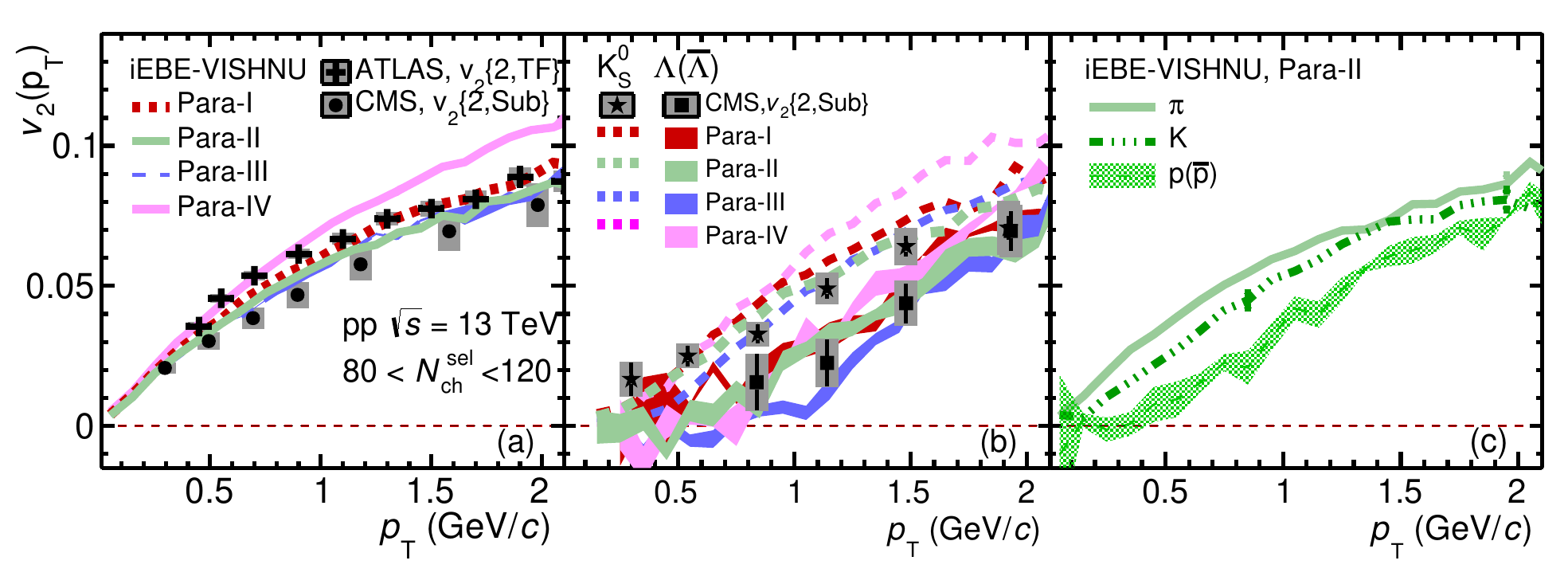}
\caption{(Color online)  $v_2(p_{\rm T})$ for all charged hadrons (a), for $K_S^0$ and $\Lambda$ (b), and for pions, kaons and protons (c) in high multiplicity pp collisions at 13 TeV, calculated by \vishnu\ with  \hijing\  initial condition. The CMS and ATLAS  data are taken from~\cite{Khachatryan:2016txc} and~\cite{Aaboud:2016yar}, respectively. }
\label{fig:vnpt-PID}
\end{center}
\end{figure*}

With the four sets of parameters listed in Table~\ref{tb:para}, we calculate the integrated flow harmonics $v_2$, $v_3$ and $v_4$ in pp collisions at 13 TeV, using \vishnu\ model with \hijing\ initial conditions. The 2-particle correlation method with a pseudorapidity gap $|\Delta\eta| > 0$,  kinematic cuts  $0.3 < p_{\rm T} < 3.0 $ GeV/$c$ and $|\eta|<2.4$ is applied in our calculations~\footnote{Following~\cite{ATLAS:2017tqk}, we firstly cut the multiplicity class with the number of all charged hadrons $N^{Sel}_{ch}$ within $0.3< p_T <3.0$ GeV and $|\eta|<2.4$, and then calculate the 2-particle cumulant $c_2\{2\}$  (as well as the following 4- particle cumulant $c_2\{4\}$) with the standard method for each unit $N^{Sel}_{ch}$ bin, which eliminate the multiplicity fluctuations. For each $N^{Sel}_{ch}$, we then map it to the average number of reconstructed charged hadrons $N_{ch}$ with $p_T>0.4$ GeV and $|\eta|<2.4$ to compare with the experimental data.}. The results and the comparisons
to the experimental data are shown in  Fig.~\ref{fig:v22}.
The colourful lines are the  results obtained with four sets of parameters (I, II, III, IV). The solid circles and triangles represent the CMS measurements of $v_2\{2\}$ and $v_3\{2\}$ with ``peripheral subtraction'' method~\cite{Khachatryan:2016txc}, the solid crosses and stars are the ATLAS $v_2\{2\}$ measurements with ``template fit''  and ``peripheral subtraction'' methods~\cite{Aaboud:2016yar}.
As shown in Fig.~\ref{fig:v22}, our model calculations reproduce the multiplicity dependence of the integrated $v_2\{2\}$ from low multiplicity ($\sim$ 30) to high multiplicity ($\sim$ 160). More specifically, our calculations with parameter sets I, II and III fit the CMS and ATLAS measurements with the ``peripheral subtraction'' method, and the one with para-IV  describes the ATLAS data from the ``template fit'' method. For low multiplicity $N_{ch} \sim$ 30, the  \vishnu\ calculations
fail to describe the CMS or ATLAS measurements with  the ``peripheral subtraction'' method, which do not significantly decrease as the data.
Figure~\ref{fig:v22} also compares $v_3\{2\}$  from \vishnu\ and from CMS measured with the ``peripheral subtraction'' method~\cite{Khachatryan:2016txc}. Our calculations with  III roughly reproduce the CMS data, while the results from para-I,  para-II and IV obviously over-estimate the data. It thus shows that the $v_3\{2\}$ measurements could further constrain the initial conditions. We also predict $v_4\{2\}$ as a function of multiplicity. Currently, the related experimental data is not publicly available, but can be compared with our model calculations in the near future~\cite{Sirunyan:2017uyl}.

In Fig.~\ref{fig:vnpt-PID}, we calculate the $p_{\rm T}$-differential $v_2(p_{\rm T})$ of all charged and identified hadrons, using \vishnu\ simulations and with the 2-particle cumulant method with a pseudorapidity gap $|\Delta\eta|>0$. Panel (a) shows a comparison between model and data for all charged hadrons, where the data from CMS and ATLAS are measured by the ``peripheral subtraction'' method~\cite{Khachatryan:2016txc} and ``template fit method''~\cite{Aaboud:2016yar}, respectively.
For para-I, para-II and III, our calculations roughly describe the CMS and ATLAS measurements within $p_{\rm T}<$ 2.0 GeV/$c$. In contrast, the calculations with para-IV slightly over predict the data above 1.0 GeV/$c$.


Fig.~\ref{fig:vnpt-PID} (b) and (c) show the $v_2(p_{\rm T})$ for $K_S^0$ and $\Lambda$  and for $\pi$, $K$ and $p$ for the multiplicity range 80 $< N_{\rm ch}^{sel} < $ 120. Clear mass orderings between $K_S^0$ and $\Lambda$ and among $\pi$, $K$ and $p$ are seen in our \vishnu\ calculations. In the hydrodynamic language, the radial flow blue-shifts the lower-$p_{\rm T}$ to higher $p_{\rm T}$ with the mass-dependent effects,  which leads to the observed  mass ordering among various hadron species.
Panel (b) also shows that the $v_{2}$ mass splitting between $K_S^0$ and $\Lambda$ is more significant for the calculations with para-III, which indicates a stronger radial flow. This consists with the results of $p_{\rm T}$-spectra in Fig.~\ref{fig:pt_spectra}, which shows that $p_{\rm T}$-spectra from para-III are flatter than other ones due to the larger radial flow. 

 \subsection{4-particle cumulant}

  \begin{figure}[thb]
\begin{center}
\includegraphics[width=0.5\textwidth]{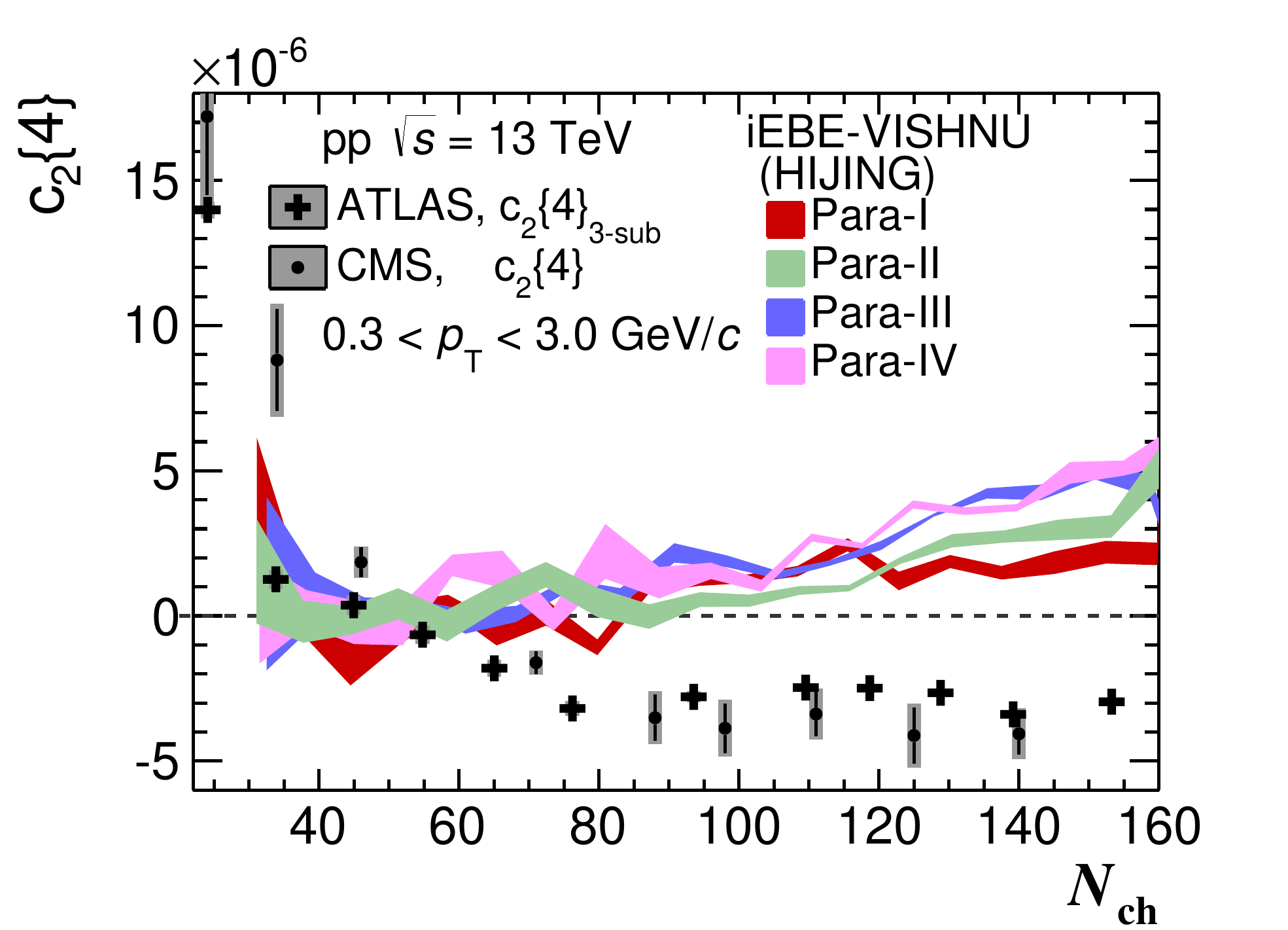}
\caption{(Color online)  $c_2\{4\}$  as a function of $N_{ch}$ for all charged hadrons in pp collisions 13 TeV, calculated by  \vishnu\ with \hijing\ initial condition using standard cumulant method. The CMS data with standard cumulant method and the ATLAS data with three-subevent method are taken from~\cite{Khachatryan:2016txc} and~\cite{ATLAS:2017tqk}, respectively.}
\label{fig:c24}
\end{center}
\end{figure}

In experiments, the observed negative $c_2\{4\}$, together with the positive $c_{2}\{6\}$ and negative $c_{2}\{8\}$,  is interpreted as a signature of collective expansion in the small systems. However, although \vishnu\ with \hijing\ initial conditions can describe the measured 2-particle correlations for both charged and identified hadrons, it fails to reproduce the negative $c_2\{4\}$  as measured by CMS and ATLAS with the standard cumulant method~\cite{Khachatryan:2016txc} and three-subevent method~\cite{ATLAS:2017tqk} ($0.3<p_{\rm T}<3.0$ GeV).
Fig.~\ref{fig:c24} shows that, for four parameter sets of \hijing\ initial conditions, \vishnu\ always predicts positive values of  $c_2\{4\}$ in the high multiplicity regime. We have also checked that these positive values are not caused
by the specific cumulant method, possible non-flow contributions or multiplicity fluctuations in
our model calculations (please also refer to the Appendix for details).

\begin{figure}[t]
\begin{center}
\includegraphics[width=0.35\textwidth]{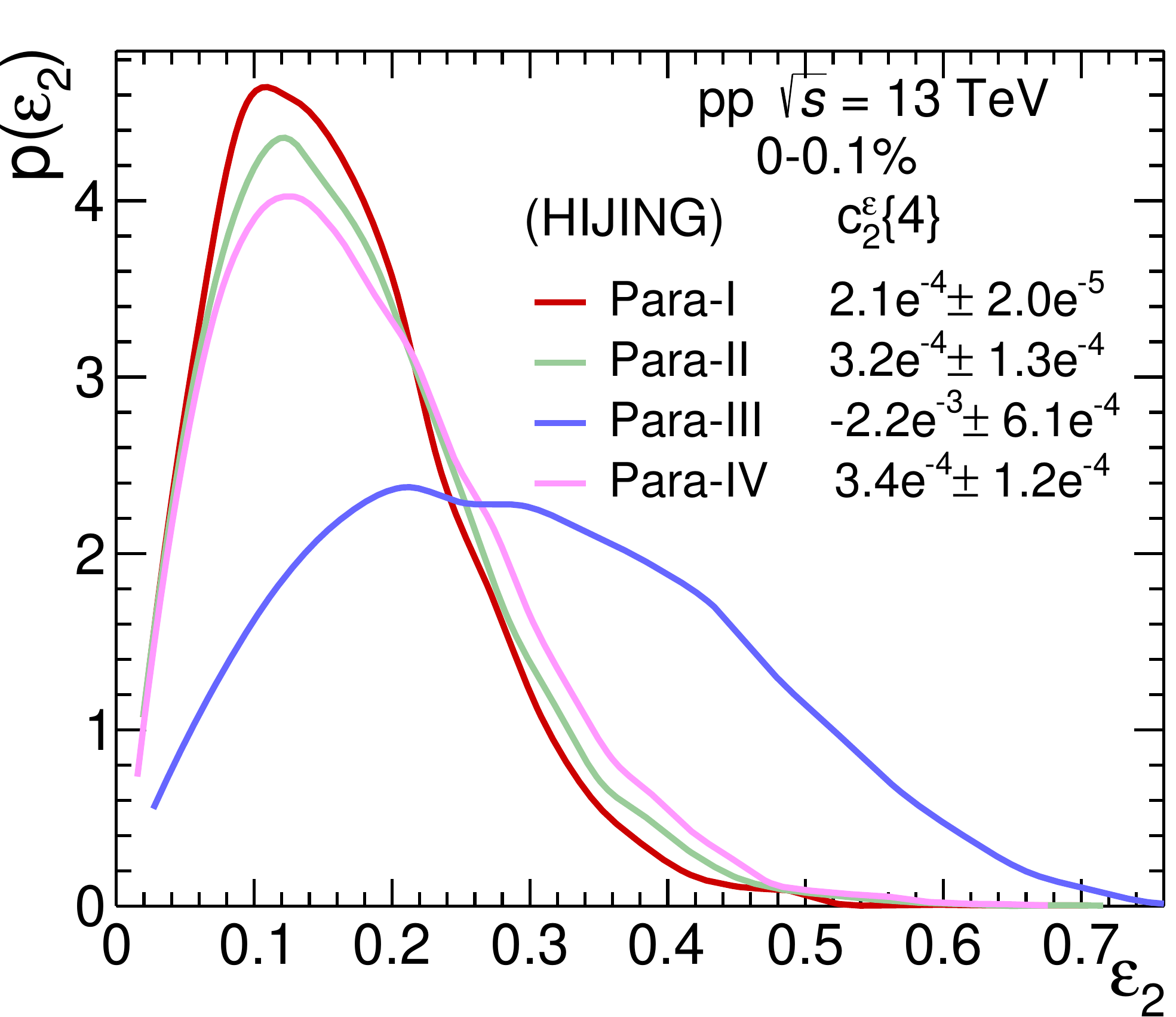}
\caption{(Color online) $\varepsilon_2$ distributions of \hijing\ initial conditions at $0-0.1\%$ centrality bin. }
\label{fig:ecc}
\end{center}
\end{figure}

In pure hydrodynamics, $c_2\{4\}=-v_2^{4}\{4\}=\left< v_{2}^4 \right>-2 \left< v_{2}^2 \right>^2$, which is influenced by both flow fluctuation and the mean value, and can be evaluated by the related $v_2$ distribution $P(v_2)$. Due to the approximate linear relationship between $v_2$ and $\varepsilon_2$,  $P(v_2)$ almost follows $P(\varepsilon_2)$ of the initial condition model~\cite{Alver:2010dn,Qiu:2011iv,Gale:2012rq}. We thus further check $P(\varepsilon_2)$ distributions of \hijing\ in Fig.~\ref{fig:ecc}, which shows that the fluctuations $\sigma_{\varepsilon_{2}}$ increases with the mean values $\left< \varepsilon_{2} \right>$. In other words, the narrower $P(\varepsilon_2)$ distribution with smaller $\sigma_{\varepsilon_{2}}$ has a smaller mean value $\left< \varepsilon_{2} \right>$, and vice versa. In Fig.~5, we also write 
the values of $c_2\{4\}^{\varepsilon}$ for the four curves. Three of them (para-I, II, IV) are positive. Correspondingly, the calculated  $c_2\{4\} $ of final emitted hadrons also present positive values as shown in Fig.~4. For para-III, $c_2\{4\}^{\varepsilon}$ has a small negative value. However, the pp fireballs in our hydrodynamic simulations do not evolve enough long time to translate that negative $c_2\{4\}^{\varepsilon}$ into an definite negative $c_2\{4\}$ as measured in experiment. For a simultaneous description of the 2- and 4- particle cumulants within the framework of hydrodynamics, other initial condition models for pp collisions should be further developed and investigated.

\section{Summary}
\label{sec:summary}

\begin{figure*}[thb]
\begin{center}
\includegraphics[width=0.9\textwidth]{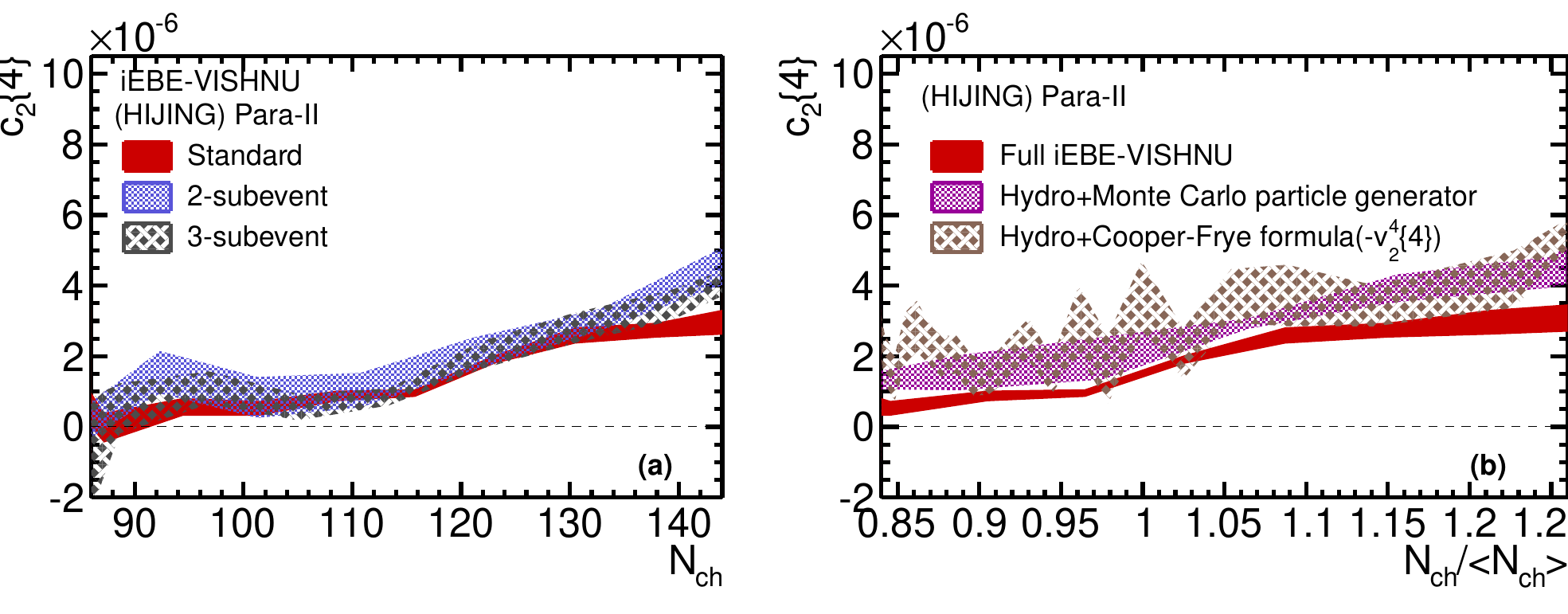}
\caption{(Color online) (a)  $c_2\{4\} $ from standard method and from 2-subevent and 3-subevent methods, calculated from \vishnu\ hybrid model with \hijing\ initial condition. (b) comparisons  of  $c_2\{4\} $ from full \vishnu\ simulations,
from hydrodynamics with Cooper-Fryer freeze-out and from hydrodynamics with Monte-Carlo particle generator.}
\label{fig:sub-event}
\end{center}
\end{figure*}

In this paper, we studied the 2- and 4-particle cumulants in proton--proton (pp) collisions at 13 TeV, using \vishnu\ hybrid model with  \hijing\ initial condition. With properly tuned parameters, our model calculations quantitatively describe the measured 2-particle cumulants, including the integrated second and third order flow coefficient $v_n\{2\}$  ($n=2, \ 3$) for all charged hadrons. In addition, \vishnu\ also reproduces the $p_{\rm T}$-differential elliptic flow $v_2(p_{\rm T})$ for all charged and identified hadrons ($K_S^0$ and $\Lambda$) in the high multiplicity pp collisions. We also predicted the  $v_2(p_{\rm T})$ of pions, kaons and protons, which shows similar characteristic mass ordering feature as the cases in Pb--Pb and p-Pb collisions, and can be further examined in experiments.

However, our \vishnu\ calculations with \hijing\ initial conditions  always give positive values of 4-particle cumulants $c_{2}\{4\}$ for various parameter sets, which can not reproduce the negative $c_{2}\{4\}$ measured by CMS and \atlas. Further investigations showed that this positive $c_{2}\{4\}$ are not caused by possible non-flow contributions, multiplicity fluctuations or the multi-particle cumulant method applied in our calculations. In fact, it is originated from the imprinted fluctuation pattern of the \hijing\ initial conditions, where the fluctuations of the eccentricity $\varepsilon_2$ increase with the increase of the mean value. Due to the approximate linear response between $v_2$ and $\varepsilon_2$, the mean value and fluctuations of the second order flow coefficient $v_2$ present a similar trend, which fails to simultaneously fit the measured 2- and 4- particle cumulants with various possible parameters. In order to simultaneously fit the flow-like data in high multiplicity pp collisions at 13 TeV within the framework of hydrodynamics, other initial condition model for the small systems should be further developed. Besides hydrodynamics, it is also necessary to investigate these 2- and 4- particle cumulants in high energy pp collisions within other theoretical approaches, to better understand the physics in high energy pp collisions.

\section{Acknowledgments}
We thank J.~Jia, M.~Zhou for providing us with the ATLAS data. We thank the discussion from J.~Jia,  U.~Heinz, X.~Zhu, W.~Li, M.~Guilbaud and M.~Zhou. WZ, HS are supported by the NSFC and the MOST under grant Nos.11435001, 11675004 and 2015CB856900. YZ are supported by the Danish Council for Independent Research, Natural Sciences, the Danish National Research Foundation (Danmarks Grundforskningsfond) and the Carlsberg Foundation (Carlsbergfondet). HX is supported by the NSFC under grand No. 11747312.

\section*{Appendix-A: $c_2\{4\}$ from standard method and 2- and 3-subevent methods}
\label{sec:Appendix}

In Ref.~\cite{Jia:2017hbm, Huo:2017nms}, it was argued that 4-particle cumulants with 2- and 3-subevent methods could further remove the residual non-flow, e.g. from the contributions of jets. Considering that non-flow effects in \vishnu\ simulations are not influenced by jet, but mainly contributed by resonance decays, we implement the standard method to calculate the 4-particle cumulant $c_2\{4\} $ in Sec.~\ref{sec:result}. In Fig.~\ref{fig:sub-event} (a), we further compare the $c_2\{4\} $  from the standard method and from the 2-subevent and 3-subevent methods, using \vishnu\ simulations  with \hijing\ initial conditions (para-I). It shows a good agreement for these three methods, which indicates that the non-flow in our calculations have been cleanly removed by the standard 4-particle cumulant, which are not necessary to further implement the 2- and 3-subevent methods that require larger statistical runs.

\section*{Appendix-B:  $c_2\{4\}$ from hydrodynamics and \vishnu\ }

To further check the positive values of $c_{2}\{4\}$ in high multiplicity ($N_{\rm ch} \sim100$) pp collisions from our model calculations, we compare $c_2\{4\}$ from (a) hydrodynamic evolution with Cooper-Fryer freeze-out, (b) hydrodynamic evolution with Monte-Carlo particle generator (c) full \vishnu\ simulations with both hydrodynamic evolution and {\tt UrQMD} afterburner. For the pure hydrodynamic calculations, we evolve the systems to $T_{sw} = $ 148 MeV and then calculate $c_2\{4\}$ with the hydrodynamic definition $c_2\{4\}=-v^{4}_2\{4\} = \left< v_{2}^4 \right> - 2 \left< v_{2}^2 \right>^2$ for case (a) and using the standard 4-particle cumulant method for cases (b) and (c). As shown in Fig.~\ref{fig:sub-event} (b), these two results from cases (a) and (b) almost overlap within error bars,  which indicates that the 4- particle cumulant $c_2\{4\}$ can properly describe the flow and flow fluctuations in a small collision system with $N_{\rm ch} \sim100$. In Fig.~\ref{fig:sub-event} (b), we also study the effects of hadronic evolution on $c_2\{4\}$ through comparing case (b) the pure hydrodynamic calculations and case (c) the full \vishnu\ simulations, which shows that the hadronic scatterings and decays slightly decrease $c_2\{4\}$.

%

\bibliography{bibliography}
%
\end{document}